# An automated, high-resolution phenotypic assay for adult *Brugia malayi* and microfilaria

Upender Kalwa[1]†, Yunsoo Park[1]†, Michael J. Kimber[2], Santosh Pandey[1]*

**Affiliations:**

[1]Department of Electrical and Computer Engineering, College of Engineering, Iowa State University; Ames, Iowa, USA

[2]Department of Biomedical Sciences, College of Veterinary Medicine, Iowa State University; Ames, Iowa, USA

*Corresponding author. Email: pandey@iastate.edu.

†These authors contributed equally to this work.

**Abstract:** *Brugia malayi* are thread-like parasitic worms and one of the etiological agents of Lymphatic filariasis (LF). Existing anthelmintic drugs to treat LF are effective in reducing the larval microfilaria (mf) counts in human bloodstream but are less effective on adult parasites. To test potential drug candidates, we report a multi-parameter phenotypic assay based on tracking the motility of adult *B. malayi* and mf *in vitro*. For adult *B. malayi*, motility is characterized by the centroid velocity, path curvature, angular velocity, eccentricity, extent, and Euler Number. These parameters are evaluated in experiments with three anthelmintic drugs. For *B. malayi* mf, motility is extracted from the evolving body skeleton to yield positional data and bending angles at 74 key point. We achieved high-fidelity tracking of complex worm postures (self-occlusions, omega turns, body bending, and reversals) while providing a visual representation of pose estimates and behavioral attributes in both space and time scales.

**Introduction:** Lymphatic filariasis (LF) is a Neglected Tropical Disease (NTD) transmitted by the bites of infected mosquitoes *(1)*. The parasites causing LF are thread-like nematodes (or worms) from the family Filarioidea – *Brugia malayi*, *B. timori*, and *Wucheria bancrofi (2)*. When a mosquito bites a human for its blood meal, the infective worm larvae enter the human skin and migrate to the lymphatic vessels to grow into adult worms. Adult worms can live for around 6–8 years, continuously shedding millions of microfilariae (mf) which circulate into host's blood and are transmitted to other mosquitoes during subsequent blood meals. Infections in humans can be acquired during childhood, causing hidden damage to their lymphatic vessels and eventually leading to painful and disfiguring disease conditions (such as lymphoedema, elephantiasis, and hydrocele). Besides the risks of permanent disability, the disease can cause significant mental stress, social stigma, poverty, and resulting economic losses to the affected communities *(1,3)*. According to the World Health Organization (WHO), LF affects over 51.4 million people in 72 countries, particularly in tropical regions of the world *(3)*.

WHO recommends the strategy of preventive chemotherapy to interrupt the transmission of LF through Mass Drug Administration (MDA) where an annual dose of drugs is given to the entire at-risk population *(1,4)*. The available drugs can reduce the number of *B. malayi* mf in the human bloodstream, which helps to curb the disease transmission by mosquitoes; however, they are less effective on adult *B. malayi (5–7)*. The specific MDA regimen to prevent LF combines three anthelmintic drugs – albendazole, ivermectin, and diethylcarbamazine citrate *(1)*. Albendazole prevents worm larvae from growing and multiplying in the human body *(8)*. It prevents the glucose uptake by worms and depletes their stored glycogen, leading to reduced cell motility, immobilization, paralysis, and eventual death of the worms. Ivermectin is an antiparasitic drug that acts on the nerves and muscles of worms *(9–11)*. It binds to glutamate-gated chloride channels, opening the channels and allowing an influx of chloride ions that hyperpolarizes the cell membrane, leading to immobilization, paralysis, and eventual death of the worms. Fenbendazole is another broad-spectrum antiparasitic drug to treat roundworms, hookworms, tapeworms, and other gastrointestinal parasites *(12)*. It selectively binds to the tubulins and disrupts the formation or depolymerization of microtubulins in worms, thereby disrupting the equilibrium of tubulins and microtubulins and leading to worm death.

While the therapeutic mechanism of action of anthelmintic drugs is still unclear, *in vitro* assays have been useful to test their effectiveness against parasitic worms. Commonly used *in vitro* assays in parasitology include the egg hatch assay, larval development assay, larval motility assay, and larval migration inhibition assay *(13–16)*. The egg hatch assay and larval development assay have been used to screen compounds against free-living *C. elegans* and plant-parasitic nematodes *(16-18)*. but are not suitable for *B. malayi* because of the requirement of a mammalian host to complete their life cycle. Both larval migration inhibition assay and larval motility assay measure the motility of parasitic worms. In larval migration inhibition assay, the ability of the larvae to pass through a fine mesh is measured where a hit compound reduces the larvae ability to migrate through the mesh. The larval motility assay is based on scoring the larvae motility over a time duration and calculating the percentage of non-motile worms in test compounds relative to control *(19)*. The earlier scoring methods used in larval motility assays were based on visual observations, and hence were subjective and dependent on the observer's definition of motility. As an example, in an observer-led study of *Haemonchus contortus*, motile worms were defined as those exhibiting normal thrashing while non-motile worms were those showing restricted or no movement *(19)*. Another visual-scoring study quantified the motility of *B. pahangi* in test compounds as normal, slight motility defect, barely moving, twitching only, and dead *(20)*. With the advent of high-resolution video capture and image processing techniques, automated motility assays for parasitic worms were developed *(21–24)*. Notable examples are the WormAssay and Worminator technologies. The WormAssay captured videos of adult *B. malayi* in 24-well plates to calculate the net pixel displacement of every worm between successive image frames and estimate the percentage of mobility reduction in drug environments *(25)*. The Worminator improved the throughput of the WormAssay to allow the user to record the motility of parasitic nematodes in 96– or 384–well plates *(26–29)*. Another study tracked the centroid location of *B. malayi* using a motorized x-y microscopy stage to quantify the worm speed, thrashing and migratory behavior in multicellular, microfluidic assays *(5)*. Researchers have also adopted the commercial real-time cell monitoring platform (xCELLigence) to measure the motility index of parasitic nematodes at real-time based on the standard deviation (over 800 points) of the cell index (CI) difference (computed as the rolling average over 20 data points) *(30)*.

The above literature indicates that motility is an informative measurement parameter in *B. malayi* phenotypic assays. However, compared to free-living nematodes, the motility of *B. malayi* is more complex and difficult to quantify at high resolution and high dimensionality. In addition, there are distinct differences between the movement patterns of adult and mf *B. malayi*, and in the repertoire of body shapes exhibited in different chemical environments. To get a holistic understanding of worm behavior and to unravel subtle phenotypes which can be overlooked, it is necessary to quantify the spatiotemporal changes in their motility using multiple parameters that capture the complex behavioral responses over time. This can significantly improve the resolution of current *B. malayi* phenotypic assays based on single-parameter measurements of the aggregate pixel displacement, mean centroid velocity or survivability, thereby revealing subtle phenotypes and undiscovered correlations amongst behavioral parameters for drug screening applications. With this goal, we developed a multi-parameter phenotypic assay (called the 'BrugiaTracker') to track the motility of adult *B. malayi* and mf. Here motility was described by spatiotemporal changes in the body shape phenotypes. We identified separate methods to describe the body shape phenotypes of adult worms and mf, as follows: (i) For adult *B. malayi*, the body shape phenotypes were tracked using six parameters – centroid velocity, path curvature, angular velocity, eccentricity, extent, and Euler Number. These parameters captured the phenotypic changes in adult *B. malayi* induced by three anthelmintics – albendazole, fenbendazole, and ivermectin – to obtain the dose response characteristics from each parameter. (ii) For *B. malayi* mf, the body shape phenotypes were tracked using the skeleton along the midline (composed of evenly distributed 74 key points from the head to the tail location). The body skeleton was iteratively estimated through successive image frames of each recorded video. Thereafter, the positional data and bending angles at each key point were calculated to estimate the number of bends along the body and the velocities at head, centroid, and tail locations. The method of iteratively extracting key point information enabled us to achieve high-fidelity tracking of complex movements, including their self-occlusions, omega turns, body bending, and reversals. The software program is automated in batch processing a set of recorded videos to produce all the parametric data in Microsoft Excel for custom analyses.

**Results:** Adult *B. malayi* videos were recorded following treatment with different concentrations of three anthelmintic drugs – ivermectin, albendazole, and fenbendazole. As shown in Fig. 1, the body shape phenotypes were diverse in different chemical microenvironments. In control and lower drug concentrations, the worm bodies were more active and convoluted with distinct knots. With increasing drug concentrations, their bodies appear to be relaxed and less convoluted with minimal knots.

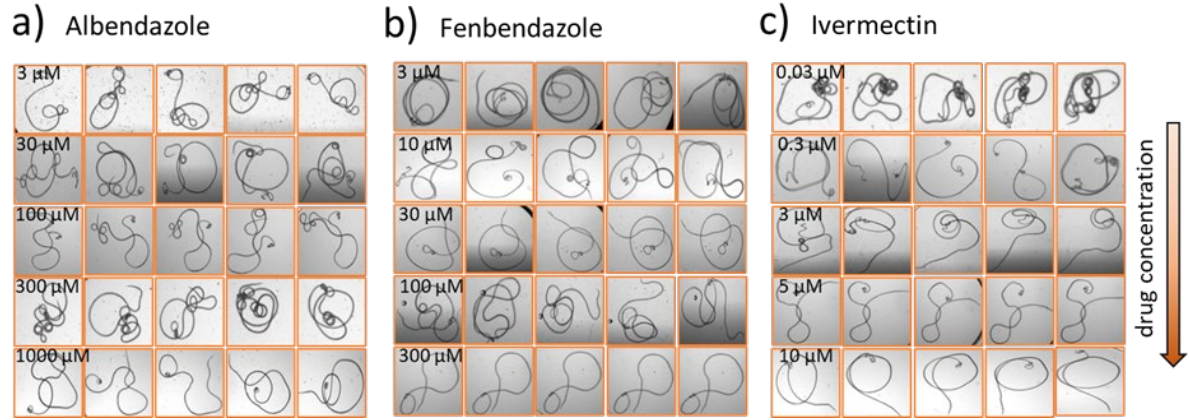

**Fig. 1.** Body shape phenotypes of adult *B. malayi* in three drugs – albendazole (a), fenbendazole (b), and ivermectin (c). Each row shows representative images of the worm in a fixed drug concentration. The drug concentrations increase from top row (lowest) to the bottom row (highest) as labelled for each row.

We quantified the body shape phenotypes indicative of the drug effects on adult *B. malayi* using six parameters as shown in Fig. 2: (i) The 'centroid velocity' was calculated as the change in body's centroid positional coordinates divided by the time difference between successive image frames. (ii) The 'path curvature' was calculated as the Menger curvature *(31)* for sets of three centroid positional coordinates. (iii) The 'eccentricity' was calculated by fitting an ellipse on the worm's body, estimating the lengths of the ellipse's major and minor axis, and thereafter calculating the eccentricity value. (iv) The 'angular velocity' was calculated from the change in angular orientation of the fitting ellipse divided by the time difference between successive image frames. (v) The 'extent' was calculated as the ratio of the worm's body area to the bounding box area. (vi) The 'Euler Number' was calculated as the number of connected components in the worm's body minus the number of holes in the image frame. Thereafter, the 'rate of extent' was calculated as the change in extent values divided by the time difference between successive image frames. Similarly, the rate of eccentricity and rate of Euler number between successive image frames were calculated. All the calculations were performed automatically by batch processing the recorded videos using the software program and outputting the data in Microsoft Excel.

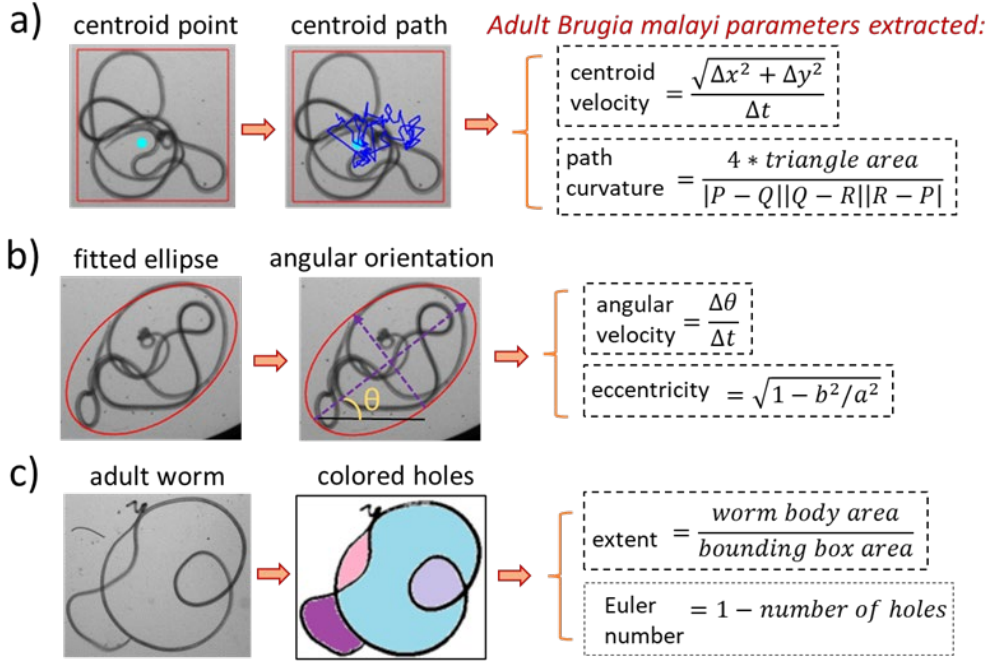

**Fig. 2.** Parameters to characterize the body shape phenotypes of adult *B. malayi*. (a) Centroid velocity is the positional change in body's centroid point (marked with a light blue dot) between successive image frames divided by the corresponding time difference. Path curvature is the Menger curvature of the path created by the centroid point (shown in dark blue) using the formula with three discrete points at a time (i.e., P, Q, R) *(31)*. (b) Angular velocity is the change in angular orientation of the fitting ellipse divided by the corresponding time difference between successive image frames. Eccentricity is calculated by fitting an ellipse on the worm body, finding the lengths of the ellipse's major and minor axis (i.e., a and b respectively), and then computing the eccentricity using the formula listed above. (c) Extent is the ratio of the worm body area to the bounding box area. Euler Number is the number of connected components in the worm body (i.e., 1 in this case) minus the number of holes in the image frame.

Figure 3 shows the raster plots of the six parameters for adult *B. malayi* in control and three drug environments. Each raster plot consists of a two-dimensional array of pixels where the color intensity of each pixel is related to the instantaneous magnitude of the parameter, and each row of pixels corresponds to one 60-second recorded video. In the raster plots, drug concentrations are labelled, and the replicates are plotted as consecutive rows of pixels. The worm motility decreased with increasing drug concentration, which was captured by the decreasing centroid velocity. Similar trends were observed for the angular velocity, rate of eccentricity, rate of extent, and rate of Euler number. On the contrary, the path curvature increased at higher drug concentrations.

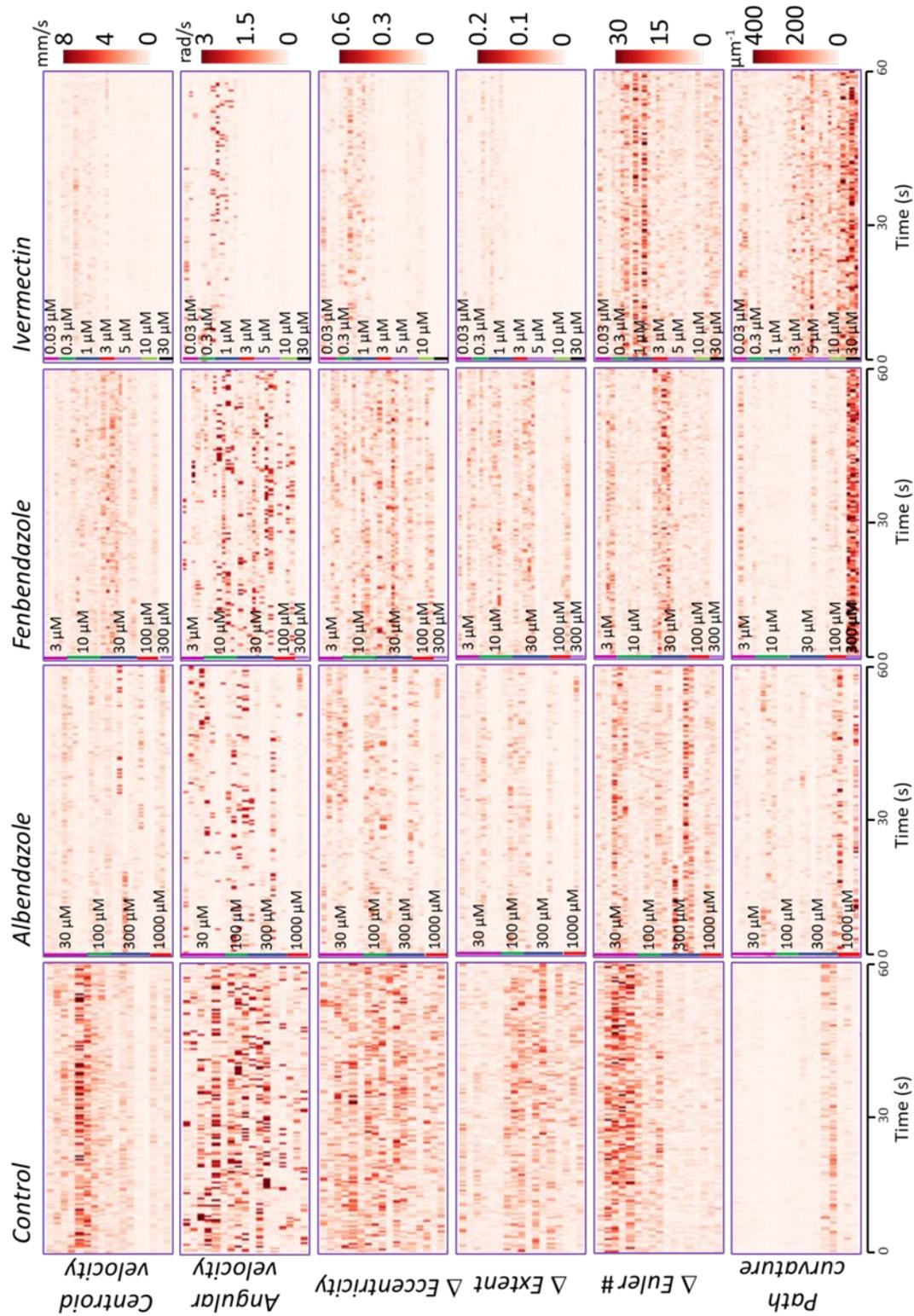

**Fig. 3.** Raster plots of the six parameters describing the body shape phenotypes of adult *B. malayi* in the control and three drug environments. The parameters are defined in Fig. 2. Each raster plot consists of a two-dimensional array of pixels where every row of pixels corresponds to one video plotted for 60 seconds and the pixel color intensity corresponds to its relative magnitude. The drug concentrations are marked, and replicates are plotted in consecutive rows.

Figure 4 plots the dose response of adult *B. malayi* in the three drug environments using the six parameters. The dose response curves were generated using the nonlinear regression model available in the GraphPad Prism 7 software. At lower ivermectin concentrations, the worms exhibited higher motility – a term called 'hyper-motility' as also observed in a previous study with adult *B. malayi (26)*.

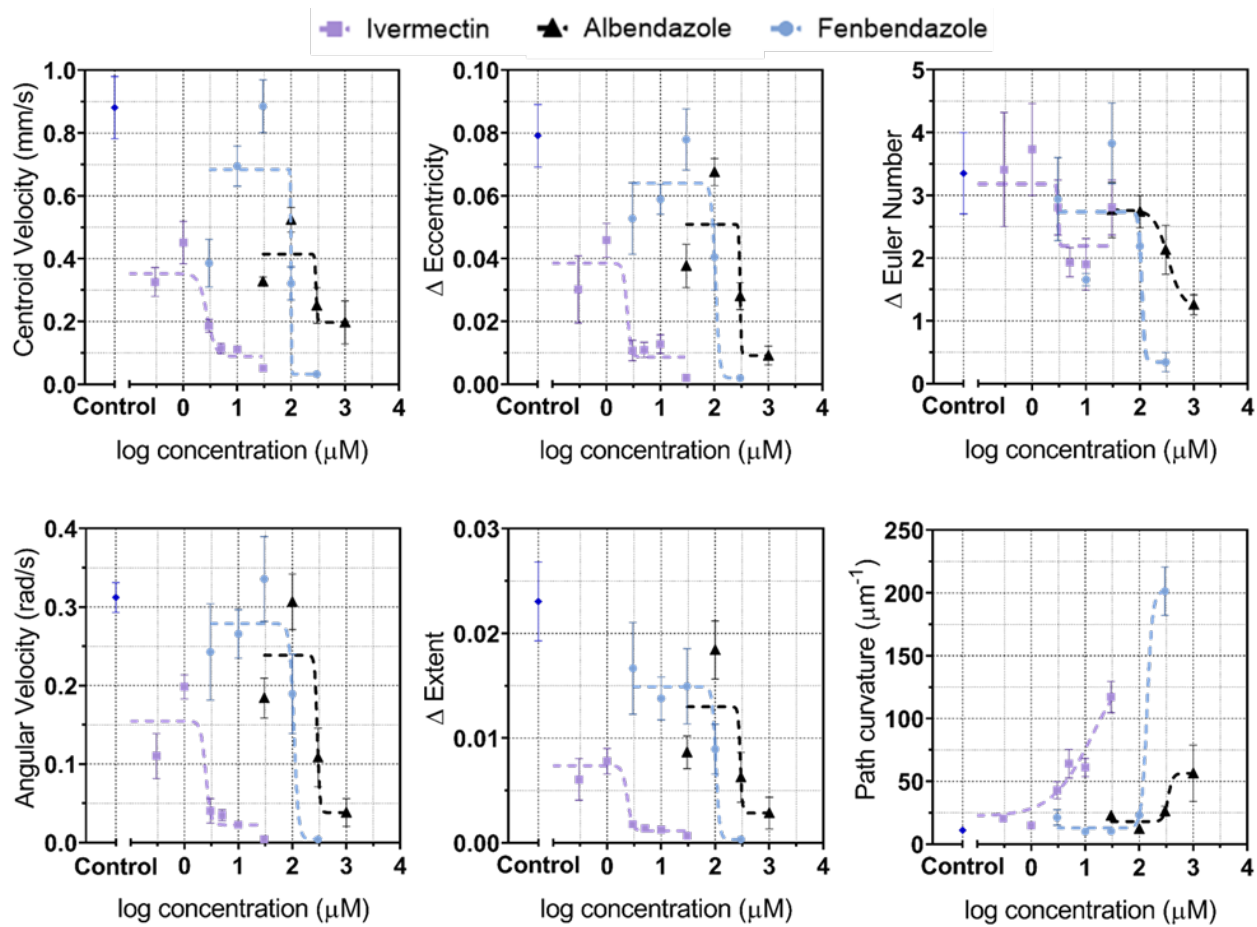

**Fig. 4.** The dose response of adult *B. malayi* in the three drug environments are plotted using the six parameters described in Fig. 2. The corresponding values in control experiments are marked. The IC$_{50}$ values are listed in Table 1.

Table 1 lists the corresponding IC$_{50}$ values (i.e., the drug concentration where the behavioral response was reduced by 50% compared to control). Excluding the path curvature, the IC$_{50}$ value for the other five parameters was between 2.3 µM to 3.04 µM for ivermectin, 290.3 µM to 333.2 µM for albendazole, and 99 µM to 108.1 µM for fenbendazole. Ivermectin was the most potent amongst the three anthelmintics, while albendazole was the least potent as observed from the dose response of the five parameters. The nature of the dose response curves shows little change over a wide range of lower concentrations, while exhibiting a steep slope beyond the IC$_{50}$ values. The path curvature increased with increasing drug concentration, and the $IC_{50}$ inferred from the path curvature was higher than those inferred from other parameters (i.e., 13.39 µM for ivermectin, 342.5 µM for albendazole, and 139.1 µM for fenbendazole). As a parameter though, path curvature is useful to study chemotactic behavior of nematodes as previously demonstrated in *C. elegans* behavioral assays *(16,32)*. The $IC_{50}$ values obtained here were in good agreement with those reported in the literature. For example, the Worminator assay reported the $IC_{50}$ value for adult *B. malayi* to be 2.22 µM for ivermectin *(26)*. Using whole-plate imaging and quantifying the percentage movement inhibition by the Lucas-Kande Optical flow algorithm, the WormAssay reported the IC50 values as 2.7 µM for ivermectin, 236.2 µM for albendazole, and 54.5 µM for fenbendazole *(25)*.

| Extracted parameters for adult Brugia malayi | Estimated $IC_{50}$ (µM) | | |
|---|---|---|---|
| | Fenbendazole | Ivermectin | Albendazole |
| Centroid Velocity | 99.73 | 2.768 | 291 |
| Angular Velocity | 107.3 | 2.468 | 290.3 |
| Δ Eccentricity | 104.3 | 2.408 | 298.6 |
| Δ Extent | 102.9 | 2.368 | 291.1 |
| Δ Euler Number | 108.1 | 3.04 | 333.2 |
| Path Curvature | 139.1 | 13.39 | 342.5 |

**Table 1.** $IC_{50}$ values of adult *B. malayi* in the three drug environments (fenbendazole, ivermectin, and albendazole).

To track the body shape phenotypes of *B. malayi* mf, the software program employed a series of image processing steps. First, the worm boundary, along with its head and tail, were identified in each image frame. The head and tail locations were differentiated using the fact that the tail forms a sharper angle than the head at the corners of the worm boundary *(33)*. Thereafter, 74 key points were denoted along the worm body and the skeleton was constructed along its midline. To automatically track the key points through all the image frames of a video, the worm boundary in the current image frame was compared with the skeleton in the previous image frame (Fig. 5a). The displacement in the worm boundary relative to the previous skeleton was calculated, and the new skeleton was constructed by adjusting the coordinates of the key points along the worm body. Knowing the new skeleton in the current image frame, the forward or reverse direction of head movement was computed by comparing the relative positions of first three key points at the head in the current skeleton to those in the previous skeleton (Fig. 5b). The worm boundary and the key points were tracked over all the image frames of recorded videos. As shown in Fig. 5c, the mf displayed a diverse range of body shape phenotypes, including body bends, omega turns and self-occlusions. The BrugiaTracker was successfully able to track the worm boundary through these scenarios.

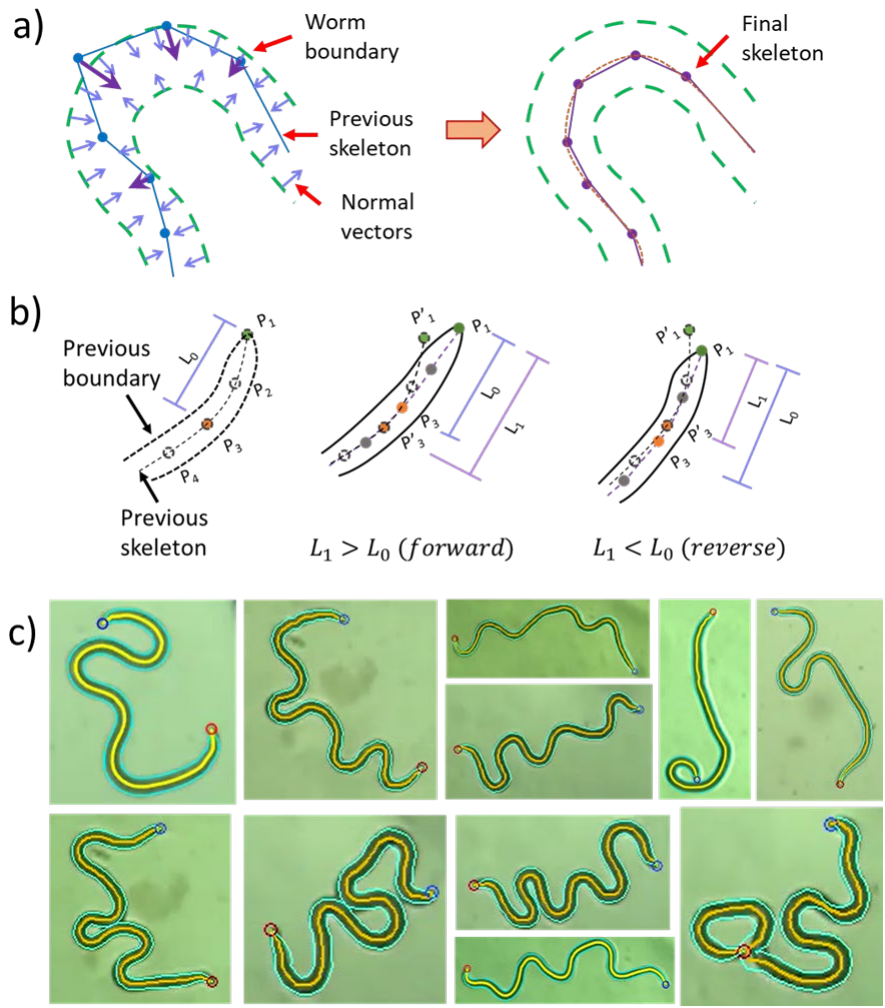

**Fig. 5.** Estimating the key points for *B. malayi* mf in drug-free environment. (a) The worm boundary was identified in the current image frame. Inward-facing vectors were drawn along the worm boundary (marked with light blue arrows) and compared with the skeletal key points from the previous image frame. The closest points between the skeletal key points and the boundary vectors were tracked along the body to produce the displacement vectors (marked as purple arrows). The previous skeletal key points were moved in the direction of their respective displacement vectors to generate the new skeleton in the current image frame where the key points follow the midline of the worm boundary. (b) The forward or reverse direction of the head was estimated by comparing the first three key points in the current skeleton ($P_1$, $P_2$, and $P_3$) with those in the previous skeleton ($P'_1$, $P'_2$, and $P'_3$). During forward worm movement, the distance $L_0$

between $P_1$ and $P_3$ was shorter than the distance L1 between $P_1$ and $P'_3$. During reverse worm movement, distance $L_0$ was greater than distance $L_1$. (c) Representative images of the *B. malayi* mf demonstrate their complex body shape phenotypes. The BrugiaTracker automatically identified the worm boundary (green border around the worm's body), the head location (blue circle), and the tail location (red circle).

Figure 6 sketches the body skeleton of *B. malayi* mf from every image frame of the recorded videos in drug-free environment. The relative percentage of forward/reverse head movement is labelled for each video, along with line bars denoting the time instances of forward/reverse head movement. In general, the body skeleton was either straighter with fewer small-sized bends (resulting in greater body displacement and greater percentage of forward velocity, Fig. 6b,e) or curvilinear with more small-sized bends (resulting in lower body displacement and lower percentage of forward velocity, Fig. 6c,f). We did not notice any significant sharp turns with body reversals which have been routinely observed in previous *C. elegans* behavioral assays *(33–36)*.

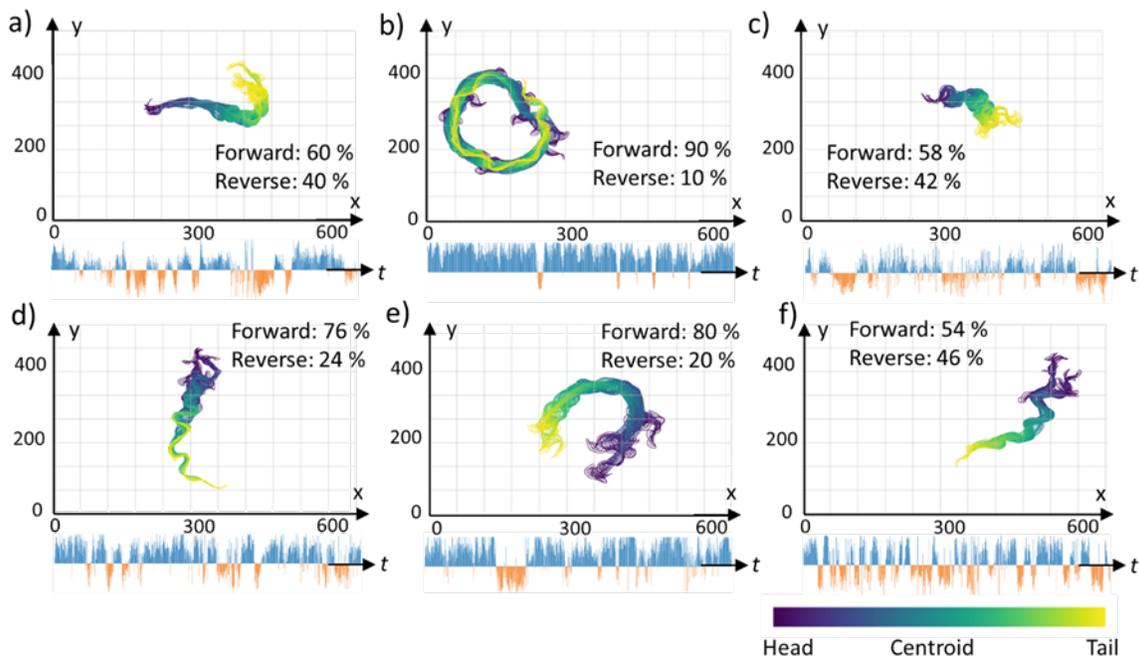

**Fig. 6.** Tracking the body skeleton and forward/reverse head movements of *B. malayi* mf on agar surface. Each x-y plot shows the body skeleton in positional coordinates. The time instances of forward and reverse movement in the six recorded videos are marked as blue and orange line bars below each plot. The 74 key points on the skeleton are color coded from head (purple) to tail (yellow).

Figure 7 shows the raster plots of the number of bends and velocity of *B. malayi* mf at the head, centroid, and tail locations. The number of bends was calculated as the signed change in curvature along the body. In contrast to *C. elegans (33–37)*, the bending frequency of *B. malayi* mf appears chaotic and there were between 6 to 17 bends for the recorded time duration. From our observations, a bend generally originated at the head region and propagated along the body to the tail region. As shown in Fig. 7b, the head velocity was relatively higher compared to the velocities of the centroid or tail locations. This may reflect consistent probing action of the head region, similar to the case with *C. elegans* where the head guides the active sensing and motor behavior response *(38,39)*.

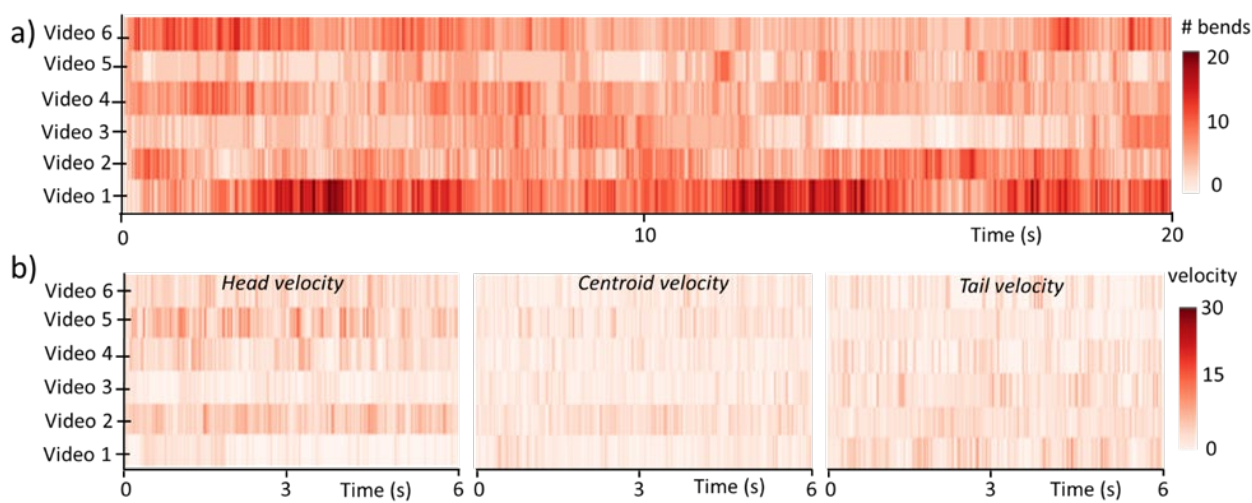

**Fig. 7.** Raster plots of the number of bends and velocity of *B. malayi* mf in buffer solution. (a) Number of body bends is plotted for six videos. (b) The velocity of the three key points at the head, centroid, and tail location are plotted.

**Discussion:** There is continued interest to screen new filaricidal drugs at the whole-organism level *(40–43)*. This requires high-resolution, high-content phenotypic assays for parasitic nematodes which are cost-effective and easy to use. In this work, our improvement in resolution and information content was achieved by recording zoomed-in videos of adult *B. malayi* and quantifying the behavioral attributes using multiple parameters – centroid velocity, path curvature, angular velocity, eccentricity, extent, and Euler Number. These parameters were selected to capture the diverse range of body shapes exhibited in different chemical environments (Movie S1-S8). Interestingly, five parameters (besides the path curvature) yielded the dose response curve with $IC_{50}$ values that were in good agreement with one another. The improved information content in *B. malayi* mf experiments was achieved by tracking the positional coordinates of the 74 key points and the body skeleton over time. This segmental data was used to gather information about the bending angles, number of bends, and velocities at the head, tail, and centroid locations. A single-parameter approach of motility scoring may not capture the diverse phenotypes in worm behavior, thus making it difficult to identify the right phenotype beforehand in drug screening experiments *(44)*. To our knowledge, this is the first attempt to track the phenotypes of adult *B. malayi* and mf with high resolution and high information content.

There are opportunities to customize the presented imaging platform for other applications. The software program is modular in design and customizable to allow incorporating new functions and metrics in the existing workflow. This can be applicable for tracking other parasitic nematodes displaying similar body shape phenotypes. The key point information can be used to extract other parameters of interest, such as the bending angles, magnitude and frequency of dominant bends, body curvature map, wave propagation velocity, wavelength, and frequency spectrum. These wave-related parameters have been extracted for *C. elegans* which show regular, sinusoidal movement *(45–48)* but have not been extracted for *B. malayi* where the movements are often irregular and chaotic. The multi-parameter approach can unravel subtle alterations in phenotypes for chemotaxis experiments *(32,38,49)* or anthelmintic drug screening. For instance, tracking the average centroid velocity itself may not reveal subtle changes in the migratory behavior of *B. malayi* within multicellular environments *(5,6)* but tracking the velocity of key points can reveal the probing behavior of the head. In addition, the imaging platform can serve as a powerful tool in parasitology to discover time-resolved correlations between different behavioral parameters, molecular and cellular targets, associated genotypes, and underlying disease mechanisms *(21,50)*.

There are certain limitations of the presented methods. Firstly, high throughput screening of multiple worms is challenging with our approach. Our setup was designed to image one worm at a time, and it is difficult to fit multiple worms in the viewing screen due to the tradeoff between the wide field-of-view and high resolution. We were able to conduct around 10-12 experiments (from sample loading to 1-minute imaging) per hour with the adult *B. malayi* on one microscope station, along with an additional 30 minutes to process and validate the recorded videos. We also found that the multi-worm tracking programs written for the shorter-length *C. elegans* and plant-parasitic nematodes were not able to track complex body shapes, such as self-occlusions, which are commonly observed in *B. malayi*. Secondly, the regular accessibility of *B. malayi* in laboratory settings is an experimental constraint. Filarial worms need to be retrieved from mammalian hosts and they are difficult to culture *in vitro* for longer time period *(20,29,51,52)*. Thirdly, our drug screening experiments have been tested on only three anthelmintics against the adult *B. malayi*. There is a need to validate the presented methods on other antiparasitic compounds and across different assays to fully understand the relative advantage of multi-parameter analysis methods. We also note that the drug test results from *in vitro* assays do not easily translate to *in vivo* scenarios in humans and animal models *(15,16,23,53)* because of the spectrum of unknown factors within the host, such as drug metabolism, drug accessibility, drug efficacy, and toxicity.

On a promising note, both whole-organism phenotypic assays and cell-based assays are viable options for the primary screening of new compounds against *B. malayi*, each providing complementary benefits in the anthelmintic drug discovery process for human diseases *(54)*. Commercialized multi-worm tracking imaging platforms designed for *C. elegans* motility studies (e.g., wMicroTracker SMART, WormScan, and xCELLigence) may provide suitable benchmarks and conceptual ideas to customize these platforms for longer-body helminths that show body shape phenotypes of greater complexities. Thus far, the adoption of commercial imaging platforms has been slow owing to their high costs. Economic incentives and investment opportunities in the field can drive down the price margins, while fostering broader participation amongst scientists and technology developers.

**Materials and Methods:** Individual adult *B. malayi* male worms were purchased from TRS Labs (Athens, Georgia, USA). The adult worms were stored in RPMI 1640 media (25 mM HEPES, 2 gm/L sodium bicarbonate, antibiotic/antimycotic, 5% HI FBS). Stock solutions of 30 mM and 100 mM albendazole, 30 mM fenbendazole, and 100 mM ivermectin were prepared using 100% dimethyl sulfoxide (DMSO). The stock solutions were serially diluted in RPMI media to prepare the working concentrations of fenbendazole, ivermectin, and albendazole. The working concentrations of three drugs were as follows: fenbendazole (3 µM, 10 µM, 30 µM, 100 µM, and 300 µM), ivermectin (0.03 µM, 0.3 µM, 1 µM, 3 µM, 5 µM, and 10 µM), and albendazole (30 µM, 100 µM, 300 µM, and 1000 µM). 1% (v/v) DMSO in RPMI media was used as the negative control. Each drug concentration and the controls were run for at least three replicates. All chemicals were purchased from Sigma Aldrich (St. Louis, Missouri, USA).

Adult *B. malayi* worms were stored in media within a 37°C, 5% $CO_2$ incubator (Thermo Fisher Scientific) for 24 hours prior to imaging. For video recording, the cover lids of 96-well black microplates (0.3 mL well volume, Corning, New York, USA) were used for housing individual worms. Within each circular indentation of the cover lid, 10 µL of the desired solution was pipetted with an individual worm and covered by a square glass coverslip (22 mm × 22 mm × 0.15 mm). The circular indentation was viewed under a stereo microscope for 30 seconds. Subsequently, video recording was started at 5 frames per second, 8x magnification, 696 × 520 pixel resolution, and 4 minutes time duration. The image acquisition setup consisted of a Leica M205C stereo microscope (Leica Microsystems, Deerfield Illinois, USA), a QiCAM 1394 Fast CCD digital camera and QCapture Software (QImaging, British Columbia, Canada), and a Dell desktop computer (having Intel Xeon E3, 8 GB RAM, and 500 GB memory storage). The QiCAM CCD camera was capable of capturing videos at 1.4 MegaPixels at 10 frames per second at full resolution, 1392 × 1040 pixels, and 4.65µm × 4.65µm pixel size. Initial trials were used to optimize the camera functions and video capture parameters. The video files were stored in .mp4 file format in their respective folders with an informative file name. For the *B. malayi* mf experiments, the head of infected *Aedes aegypti* mosquitoes were excised *(51)* to release the mf on glass slides coated with agarose gel film using standard fabrication techniques *(15)*.

A custom software program, BrugiaTracker, was written in MatLab to batch process the recorded videos of adult and mf *B. malayi* worms. The software program read the first image frame in every video and detected the circular well boundary using the Circle Hough Transform. The average pixel intensity was calculated for the well area. Thereafter, a user-defined percentage thresholding technique was applied to create a mask and identify the worm's body. The following body shape parameters were extracted from the adult worm's body in each image frame – centroid velocity, path curvature, angular velocity, eccentricity, extent, and Euler Number. The data analysis was performed using GraphPad Prism 7 (GraphPad Software, Boston, Massachusetts, USA) and the data was plotted using the Matplotlib library in Python (version 3.11.0). For *B. malayi* mf experiments, the BrugiaTracker program extracted the body shape parameters by detecting the worm boundary, differentiating the head and tail locations, identifying 74 key points along the worm's body, and tracking the body skeleton through the videos as shown in Fig. 5.

**Acknowledgments:** We are grateful to our students and colleagues for their help and guidance on the experiments, including Zach Njus, Hiruni Harischandra, Prof. Lyric C Bartholomay (UW), and Prof. Judy Sakanari (UCSF).

**Funding:** The project was partially supported by funding from:

U.S. National Science Foundation grant NSF IDBR-1556370 (SP)

U.S. Department of Agriculture grant 2020-67021-31964 (SP)

**Author contributions:**

Conceptualization: UK, YP, MJK, SP

Methodology: UK, YP, MJK, SP

Investigation: UK, YP, SP

Visualization: UK, YP, SP

Funding acquisition: SP

Project administration: MJK, SP

Supervision: MJK, SP

Writing – original draft: UK, YP, SP

Writing – review & editing: UK, YP, MJK, SP

**Competing interests:** Authors declare that they have no competing interests.

**Data and materials availability:** All data are available in the main text or the supplementary materials. Additional data related to this paper may be requested from the corresponding author.

## Supplementary Materials

Supplementary Text

Movies S1 to S8